\documentclass[prl,twocolumn,groupedaddress,showpacs]{revtex4}
\usepackage[english]{babel}
\usepackage{epsfig}
\usepackage{amsmath}
\usepackage{amssymb}
\usepackage{amsfonts}
\usepackage{array}
\usepackage{subfigure}
\usepackage{psfrag}

\DeclareMathOperator{\e}{e}

\begin{document}
\bibliographystyle{apsrev}

\title{Non-monotonic dependence of the rupture force in polymer chains on their lengths}

\author{S. Fugmann and I. M. Sokolov}

\affiliation{Institut f\"ur Physik,
Humboldt-Universit\"at zu Berlin,
Newtonstra\ss e 15, D-12489 Berlin, Germany}

\date{\today}

\begin{abstract}
We consider the rupture dynamics of a homopolymer chain pulled at one end 
at a constant loading rate. Our model of the breakable polymer is related
to the Rouse chain, with the only difference that the interaction between
the monomers is described by the Morse potential instead of the harmonic one, and
thus allows for mechanical failure. We show that in the experimentally relevant
domain of parameters the dependence of the most probable rupture force on
the chain length may be non-monotonic, so that the medium-length chains 
break easier than the short and the long ones. The qualitative theory of the 
effect is presented.
\end{abstract}


\maketitle

Probing the mechanical response of single molecules to an external load 
has attracted considerable interest in recent years.
In molecular failure experiments \cite{Rief97,Mehta99,Grandbois99,Cui06,Friedsam03,Neuert07} 
and experiments on protein unfolding \cite{Strunz99} a force grows linearly in time until 
a molecule breaks or changes its structure. The dynamic force
spectroscopy \cite{Evans01,Strunz99} delivering the spectrum of the 
rupture forces at different loading rates gives deep insights into the internal 
dynamics of molecules \cite{Garg95,Dudko03,Dudko06,Dias05}. In all these 
experiments polymers play an outstanding role, either as elastic couplers 
or as a subject of study \cite{Embrechts08, Kuehner06, Friedsam03}. It was 
shown, that even the mechanical properties of passive polymer spacers can 
affect the outcome of pulling experiments \cite{Friedsam03,Kuehner06,Neuert07}. 
Recently a strong impact of the polymer size on its rupture dynamics has 
been reported \cite{Embrechts08,Fugmann09}.

The overall picture of the homopolymer mechanical failure is very similar to the one
of the failure of macroscopic fibers, as represented as a sequence of links.  
Under very slow ramping of the external load each link of the chain is subjected to the 
same force, and it is mostly the microscopic mechanism of a failure of a single link,
which discriminates between the two situations. In a macroscopic case the single link
failure is due to the preexisting defects, so that the survival probability
$W_1(f)$ of a link under force $f$ typically follows a power law. 
The probability that all the $N$ links are intact is then given by 
\begin{equation}
W_{N}(f)=W_1[f(t)]^{N}
\label{eq:probN}
\end{equation}
and tends to a Weibull law for fibers long enough. Moreover, the longer fibers typically 
break at smaller forces, since the probability to find a weak link grows with the 
fibers' length, see e.g.~\cite{Padgett95}. In a microscopic homogeneous polymer chain, the 
single link breakdown is thermally activated; the breakdown probability at a given force
at a given time follows an exponential law (as obtained from the Kramers theory of
thermally activated barrier crossing), and therefore $W_{N}(f)$ tends to a Gumbel 
distribution. However, the fact that the longer chains break more easily
still holds: the larger is the number of links, the higher is the probability that 
one of them breaks \cite{Embrechts08}. At higher loading rates the situation with the polymer chain gets more involved. 
As we proceed to show, a complex interplay between the temporal aspects of thermally
activated single link breakdown and the force redistribution along the chain leads
to new features of the polymer failure problem. Thus, the behavior of the most 
probable force at breakdown as a function of the chain's length gets non-monotonic:
the medium-length chains break more easily than the short and the long ones.
The effect is not small and therefore is pertinent to experimental observation.
The breakdown force is reduced by $5-7\%$ compared to
the extremal cases of a single bond and of an infinitely long chain at the chosen values of the parameters, and the effect is expected to be the more pronounced the softer the bonds are since then the thermally activated breakdown happens earlier while the Rouse time, being a measure for the timescale of force propagation, becomes large.

In what follows we first discuss our theoretical model and present the results of
its numerical simulations. We then turn to the analytical description of
the behavior observed, giving a full qualitative picture of the effect and then
give its simple explanation.

Like in \cite{Fugmann09} our model corresponds to a chain of $N$ monomers interacting 
via the Morse potential
\begin{equation} 
U(q)=\frac{C}{2\alpha} \left(1-e^{-\alpha q}\right)^2\,,
\label{eq:morse_pot}
\end{equation} 
which parameterizes the interaction energy in terms of dissociation energy 
$C/(2\alpha)$ and stiffness $C \alpha$. This is a simple prototype 
of an intramolecular interaction potential which offers fragmentation. Otherwise, 
the model is identical to the Rouse one \cite{DoiEdwards86}: we disregard 
hydrodynamical interactions and describe the interaction of the monomers with 
the heat bath via independent white noises. The constant loading enters through 
an additional time dependent potential of the form
\begin{equation}
L(q,t)=-qRt\,, 
\label{eq:loading_pot}
\end{equation}
with loading rate $R$. The load is denoted by $F(t)=-\partial L(q,t)/\partial q$ and 
is applied at one end of the chain while the other end is fixed.

At smaller loads the overall potential has two extrema,
$q_{extr}^{\pm}(t)=\ln(2)/\alpha-\ln\left(1\pm \sqrt{1-F(t)/F_c}\right)/\alpha$, 
a minimum $q_{extr}^+$ corresponding to a metastable state of the pulled bond, 
and a maximum $q_{extr}^-$ providing the activation barrier. There exists a 
critical load $F_c=F(t_c)=C/4$ for which the extrema merge at 
$q_c=\ln(2)/\alpha$ and disappear. In the purely deterministic dynamics the 
Morse bond breaks exactly at $t_c=F_c/R$. Since the system is in contact to a 
heat bath at temperature $T$, its overdamped dynamics is described by a set 
of $N$ coupled Langevin equations
\begin{equation}
\gamma\dot{q}_i=-K(q_i-q_{i-1})+K(q_{i+1}-q_{i})+\sqrt{2k_BT\gamma}\xi_i+Rt\delta_{i,N}\,,
\label{eq:setLangevin1} 
\end{equation}
with $K(q) = -\partial U(q)/\partial q$, Gaussian white noise $\xi(t)$, 
the Boltzman constant $k_B$ , the friction coefficient $\gamma$ and $q_0\equiv0$. 
We introduce $c=C/\gamma$ ($\left[c\right]=nm/\mu s$), $r=R/\gamma$ 
($\left[r\right]=nm/\mu s^2$) and $f=F/\gamma$ (in the
following $f$ is referred to as force). The diffusion coefficient is 
denoted by $D=k_BT/\gamma$ ($\left[D\right]=nm^2/\mu s$).

The set of coupled equations (\ref{eq:setLangevin1}) is integrated by 
use of a Heun integration scheme. A chain is considered as broken as soon as
one of the reaction coordinates $q_{i+1}-q_i$ overcomes the activation barrier. 
Statistics stem from an ensemble of at least $10^3$ simulation runs.
\begin{figure}
\begin{center}
\includegraphics[height=6.cm,width=8.cm]{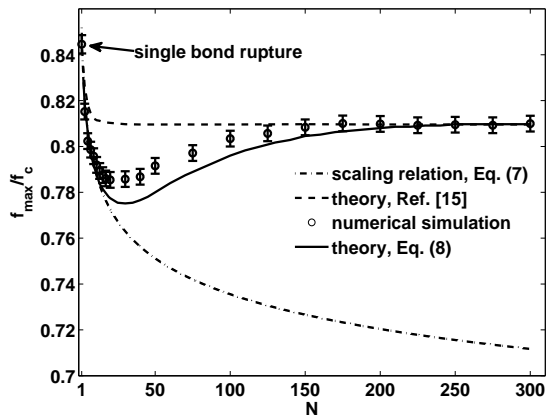}
\end{center}
\caption{\label{fig:rabe_length} Most probable rupture force $f_{max}$ as a 
function of the chain length $N$. The
parameter values are $c=3.5nm/\mu s$,
$D=2\times10^{-3}nm^2/\mu s$, $r=10^{-3}nm/\mu s^2$, and $\alpha=10nm^{-1}$. 
Error bars indicate the uncertainty due to the binning of numerical data.}
\end{figure}

In Fig.~\ref{fig:rabe_length} we present the numerically obtained most
probable rupture force $f_{max}$ as a function of the chain length $N$
for a fixed value of the loading rate $r$. For small chain lengths the 
numerically obtained most probable rupture force (symbols) follows the 
scaling relation given in Eq.~\eqref{eq:scaling_rabe_n} reaching a minimal 
value for an intermediate value of $N$. A further increase in the chain length 
results in an increase of the most probable rupture force and eventually in 
a saturation of the latter. A theoretical description following the ansatz 
in \cite{Fugmann09} (dashed line) gives in a good agreement the breakdown force
for shorter chains as well as in the saturation regime, i.e. for very long ones. 
The curve obtained from Eq.~\eqref{eq:a_pdf} of the present contribution 
(solid line) gives the correct qualitative behavior in the whole domain of the chains'
lengths.

\begin{figure}
\begin{center}
\includegraphics[height=6.cm,width=8.0cm]{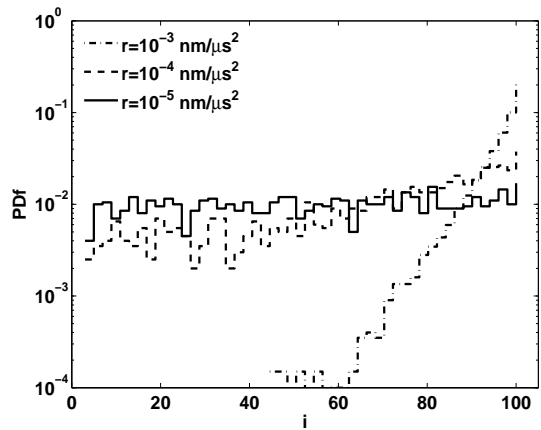}
\end{center}
\caption{\label{fig:pdf_length} Probability density of the position of the 
breakdown $i$ in a chain of $N=100$ bonds for different values of the loading 
rate $r$ (given in the legend). The remaining parameter values are the same as in Fig.~\ref{fig:rabe_length}.}
\end{figure}

The distribution of the position of the breakdown in a chain of $N=100$ links 
is shown in Fig.~\ref{fig:pdf_length}. For a small loading rate, i.e., 
$r=10^{-5} nm/\mu s^2$ (solid line), there is only a very slight decrease 
of the rupture probability density along the chain. Virtually all bonds 
contribute equally to the rupture process. The situation changes 
drastically when passing to higher loading rates (dashed and dashed-dotted lines). 
The rupture probability density decreases fast along the chain. 
For $r=10^{-3} nm/\mu s^2$ only half of the chain contributes to the rupture process.

\begin{figure}
\begin{center}
\includegraphics[height=6.cm,width=8.0cm]{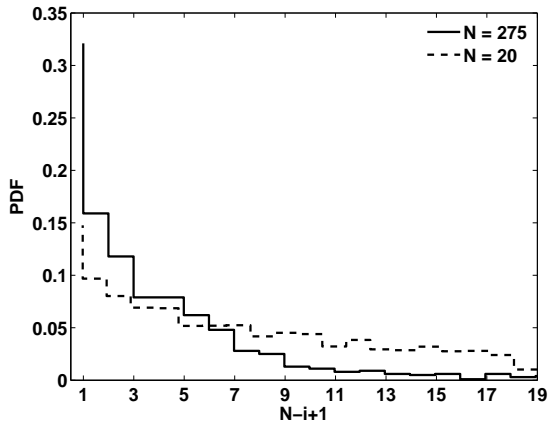}
\end{center}
\caption{\label{fig:pdf_length2} Probability density of the position of the 
breakdown $i$ in a chain of $N$ bonds for one fixed value of the loading rate, 
$r=10^{-3} nm/\mu s^2$, and two different chain lengths $N$ (given in the legend). 
The remaining parameter values are the same as in Fig.~\ref{fig:rabe_length}.}
\end{figure}

In Fig.~\ref{fig:pdf_length2} we present the distribution of the position of 
breakdown in the chain for a fixed value of the loading rate and two 
different values of the length $N$. One readily infers that the probability of a bond breakdown at a given site decays for a longer chain (solid line) faster with the distance from the pulled end than it does for a shorter one (dashed line). Thus, although the longer chains offers a larger number of possible 
breakdown sites, the number of bonds which contribute to the rupture 
process becomes smaller reaching a constant---loading rate dependent---
value in the limit of a semiinfinite system.

In order to derive an analytical description of the chain rupture process we 
first recall the model describing the single bond rupture. The probability 
$W_1(t)$ that a single breakable bond remains intact
can be expressed through the following kinetic equation
\cite{Tshiprut06,Dudko03,Dudko06,Raible04} ${dW_1(t)}/{dt}=-k(t)W_1(t),$
with $k(t)$ being the Kramers rate \cite{Kramers40,Hanggi90}. 
Taking $f(t)=rt$ we can rewrite the kinetic equation in the form
$dW_1(f)/df=-k(f)/r W_1(f)$. The measured probability density function (PDF) for the rupture forces 
$P_1(f)$ then is $P_1(f)=-dW_1(f)/df$.

Under the assumption that $f$ is close to $f_c$ when bond rupture
occurs, it is usual to expand the potential around the inflection
point $q_c$ up to the third order in deviations from $q_c$
\cite{Garg95,Dudko03,Dudko06}. We note that the breakdown properties
of the chain only depend on the behavior of the potential close to the
point of critical load, which are universal
\cite{Dudko06,Lin07,Friddle08}. The Morse potential gives
a convenient parametrization of the situation in terms of dissociation
energy and stiffness. In the limit of small loading rates the most probable
rupture force $f_{max}$ follows the scaling relation \cite{Dudko03, Fugmann09}
\begin{equation}
f_{max}=f_c\left[1-\left(\frac{\ln\left(v/r\right)}{w}\right)^{\frac{2}{3}}\right]\,,
\label{eq:scaling}
\end{equation}
with $v=c\alpha^2D/(8\pi)$ and $w=c/(3\alpha D)$.

Passing to a chain of $N$ breakable bonds we consider the rupture dynamics of 
different bonds to be independent. The probability that a bond $i\in 1\dots N$ is 
intact is denoted by $W_1(f_i(t))$. The pulling force at the chain end is $f\equiv f_N(t)$. 
The equation for the survival probability of the chain reads:
\begin{equation}
W_N(t)=\exp\left\{\int_0^N\ln\left(W_1(f(x,t))\right)dx\right\}\,.
\label{eq:wl}
\end{equation}

\begin{figure}
\begin{center}
\includegraphics[height=6.cm,width=8.cm]{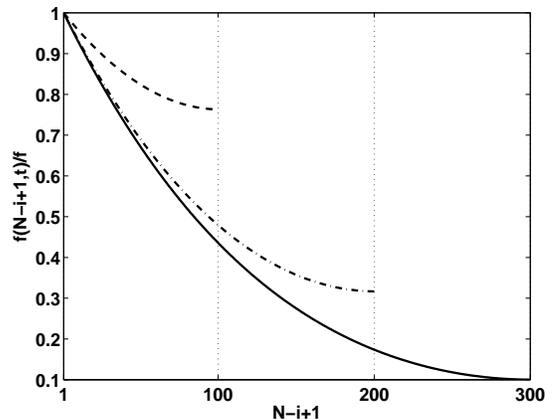}
\end{center}
\caption{\label{fig:force_profile} Force profile in a Rouse chain with a coupling constant $c\alpha$, $t=600\mu s$. The remaining parameter values are the same as in Fig.~\ref{fig:rabe_length}.}
\end{figure}

If the typical rupture time $t_{max}=f_{max}/r$ 
is much larger than the Rouse time $\tau$ of the system, $\tau=N^2/(c\alpha\pi^2)$,
(which is the case either for short chains or for small loading rates)  
each bond in the chain experiences the same force $f(x,t)=f$. 
The probability that a bond breaks in an interval $[f,f+df]$ is $P_N(f)=NW_1(f)^{N-1}P_1(f)$ 
and it is given by the same expression as $P_1$ with $v$ changed for $Nv$. 
This gives the scaling relation for the most probable 
rupture force \cite{Fugmann09}
\begin{equation}
f_{max}=f_c\left[1-\left(\frac{\ln\left(\frac{Nv}{r}\right)}{w}\right)^{\frac{2}{3}}\right]\,.
\label{eq:scaling_rabe_n}
\end{equation}
The opposite situation is more involved. First, we have to calculate $f(x,t)$.
To do this, we note that the barrier crossing events are very rare:
Most of the time the motion of the monomers takes place close to the quadratic 
potential minima. Therefore, like in Ref. \cite{Fugmann09}, the dynamics of the chain
can be approximated by a Rouse one. At difference to \cite{Fugmann09} the
boundary condition at the grafted end is explicitly taken into account. This
leads to the dependence of the force profile on the chain's length as shown in
Fig.~\ref{fig:force_profile}. 
Linearizing the force profile close to the pulled end and inserting it into 
Eq.~\eqref{eq:wl} we eventually, derive the PDF 
\begin{eqnarray}
&& P(N,t)=-W(N,t)\left\{\frac{2vf_c}{3rw^{\frac{2}{3}}}\frac{d_N(f)}{g_N(f)^2}\times\right. \label{eq:a_pdf}\\
&&\left.\left(\Gamma\left(\frac{2}{3},a(f)\right)-\Gamma\left(\frac{2}{3},a(f)\left(1+S(N,f)\right)^{\frac{3}{2}}\right)\right)\right. \nonumber \\
&&\left.-v\left[\frac{\e^{-a(f)}}{rg_N(f)}-\frac{1-d_N(f)N}{rg_N(f)}
\e^{-a(f)\left(1+S(N,f)\right)^{\frac{3}{2}}}\right]\right\}\,, \nonumber
\end{eqnarray}
with the probability function

\begin{eqnarray}
&& W_N(f) = W_0^N\exp\left[-\frac{2vf_c}{3rg_N(f)w^{\frac{2}{3}}}\right. \label{eq:a_wl2}\\
&& \left. \left(\Gamma\left(\frac{2}{3},a(f)\right)-\Gamma\left(\frac{2}{3},a(f)\left(1+S(N,f)\right)^{\frac{3}{2}}\right)
\right)\right]\,. \nonumber
\end{eqnarray}
We introduced 
\begin{eqnarray}
&& g_N(f)=\\
&&f-\frac{8c\alpha\tau}{N}\sum_{n=1}^{\infty}\sin\left(\frac{n\pi}{2}\right)\left(\frac{f}{n^2}-\frac{4r\tau\left(1-\e^{-\frac{n^2 f}{4r\tau}}\right)}{n^4}\right)\nonumber  \\
&& \times\left\{\sin\left(\frac{n\pi \left(N-\frac{1}{2}\right)}{2N}\right)-\sin\left(\frac{n\pi \left(N-\frac{3}{2}\right)}{2N}\right)\right\}\,,\nonumber
\end{eqnarray}
$S(N,f)=g_N(f)N/(f_c-f)$, $d_N(f)=dg_N(f)/df$ and $a(f)=w\left(1-f/f_c\right)^{3/2}$. $\Gamma$ is the upper incomplete Gamma function.

The analytical description given by Eq.~\eqref{eq:a_pdf} 
(solid line in Fig.~\ref{fig:rabe_length})) agrees well with the outcome of 
the numerical simulations. The nonmonotonous behavior of the rupture forces 
is nicely reproduced and in the limit of short and long chains the theory 
agrees also qualitatively. Furthermore the chain lengths which minimizes the 
most probable rupture force coincide with the ones derived from the numerical 
simulations. Deviations can result from the harmonic approximation in the 
derivation of the force profile: the decay of the force profile along 
the chain for a soft Morse potential is expected to be more pronounced than in the Rouse
chain, so that less bonds contribute to the rupture process. This might explain the shift 
of the theoretical curve to lower rupture forces compared to
numerical data points.

The overall behavior can therefore be explained as follows. 
A short chain has only a few bonds that can break. Each of them 
feels practically the same force $f$, since the forces $f_i$  
acting on the bonds decrease only slightly with their distance 
from the pulled end, see Fig.~\ref{fig:force_profile}. The longer is the chain, the more breakable 
bonds are present, however each of them is subject to the tension
which is smaller than $f$ and decays with the chain's length. 
The interplay between the thermally activated rupture of a single 
bond and the force distribution along the chain generates a non-monotonous 
behavior of the typical rupture forces.

Let us summarize our findings. Compared to the single bond breaking, 
the existence of the chain introduces new aspects into
rupture dynamics, the most important being
the delayed stress propagation along the chain. 
We show that the most probable rupture force decreases with the length of the
chain  as $f_{max}\propto -(\ln (const\,N))^{2/3}$ for short chains and saturates at the value depending on 
the loading rate for very long ones. In between it can exhibit a non monotonous behavior:
the most probable rupture force attains its minimum for a certain intermediate chain length. 
These results are obtained via direct numerical simulations of a Rouse-like model
(however with anharmonic Morse interaction potential between the monomers instead of
a harmonic one for a genuine Rouse chain) and are well reproduced by a generalization of an analytical approach
put forward in our previous study \cite{Fugmann09}. The qualitative explanation of the effect involves a complex 
interplay between the force propagation into the chain and the extreme-value statistic underlying rupture. Since the effect is rather dependent on the different timescales in the system under study than on the specific parameter values it is pertinent to experimental observation.

\begin{acknowledgments}
The authors thankfully acknowledge valuable discussions with W. Ebeling.
This research has been supported by DFG within the SFB 555 research collaboration program.
\end{acknowledgments}


\end{document}